\documentclass{jpp}
\usepackage{graphicx}

\usepackage[utf8]{inputenc}
\usepackage[T1]{fontenc}
\usepackage{amsmath}
\usepackage{makecell}
\usepackage{enumitem}
\usepackage{upgreek}
\newlist{assumptions}{enumerate}{1}

\setlist[assumptions]{label=\textbf{\alph*)},ref=\alph*}

\title{On shear Alfv{\'e}n wave-induced energetic ion transport in optimized stellarators}

\author{
A.~R.~Knyazev\aff{1}\corresp{\email{a.knyazev@columbia.edu}},
A.~Lachmann\aff{1},
A.~G.~Goodman\aff{2},
A.~S.~Hyder\aff{1},
M.~Czekanski\aff{3},
D.~Spong\aff{4}
\and E.~J.~Paul \aff{1}}

\affiliation{
\aff{1}Department of Applied Physics and Applied Mathematics, Columbia University, New York, New York 10027, USA
\aff{2}Max Planck Institute for Plasma Physics, Wendelsteinstraße 1, D-17491 Greifswald, Germany
\aff{3}Department of Statistics and Data Science, Cornell University, Ithaca, NY 14853-5169
\aff{4}Oak Ridge National Laboratory, Oak Ridge, TN 37831
}

\begin{document}

\maketitle

\begin{abstract}
In this work, we investigate prompt ion drift orbit losses caused by shear Alfv{\'e}n waves (SAW) in quasi-symmetric (QS) and quasi-isodynamic (QI) stellarators optimized for equilibrium confinement of energetic particles (EPs). We use the ideal reduced MHD model for SAW perturbations and study their impact on collisionless EP drift dynamics. We present a semi-analytical model for resonance between the passing EP and SAW, generalized to arbitrary quasi-symmetric configurations including the quasi-poloidal case relevant to QI equilibria. Analysis reveals that an increase in the number of field periods $N_{\rm fp}$ suppresses stochasticity in quasi-helical (QH) and quasi-isodynamic, but not quasi-axissymmetric (QA) stellarators. We show that wave-induced transitions between passing and trapped orbits cause significant losses in QA and QH, but not in QI configurations. For the considered equilibria at scales relevant to fusion power plants (FPPs), we numerically determine SAW amplitudes needed to induce prompt loss of fusion-born alpha particles. Using the weighted Birkhoff averaging technique, we confirm that the onset of prompt losses across all orbit classes occurs with the onset of stochasticity in ion motion. This motivates extending the stochasticity-onset criterion beyond passing orbits in future work.
\end{abstract}

\section{Introduction}
\label{sec:introduction}
Recent optimization \citep{landreman2022magnetic, goodman2023constructing} of stellarator equilibria has reduced orbit losses of fusion-born energetic ions to levels compatible with FPP design. This has motivated academic and private research \citep{hegna2025infinity, lion2025stellaris, gates2025stellarator, goodman2025aquasi} to develop a physics basis for stellarator FPPs. In addition to sufficiently low equilibrium losses, a viable FPP design needs to mitigate prompt energetic ion losses induced by shear Alfv{\'e}n wave (SAW) perturbations, which are known to produce significant transport in tokamaks (see \citet{breizman1995energetic}, \citet{sharapov2021energetic}, and references therein). Accordingly, there is significant recent interest \citep{white2022poor,paul2023fast} in assessing energetic particle losses from stellarators in the presence of shear Alfv{\'e}n waves. These initial studies demonstrated that resonant SAW perturbations can drive significant losses of fusion-born alphas in FPP designs at normalized perturbation amplitudes that are consistent with previous experiments (see \cite{toi2011energetic}, \citet{nishimura2013simulation} and references therein). Stellarator FPP designs, therefore, require an assessment of SAW stability in the presence of energetic fusion products \citep{varela2024stability}, and wave-induced transport analysis \citep{hegna2025infinity} as part of their physics baseline. Besides FPPs, sufficient confinement of energetic particles is necessary for a viable design of a beam-driven stellarator neutron source like the recently proposed EOS device \citep{swanson2025scoping}.

Previous SAW-induced transport studies \citep{white2022poor,paul2023fast} considered optimized stellarator equilibria and identified stochastic ion transport as a significant loss channel. Resonance overlap analysis for precisely quasi-axissymmetric (QA) and quasi-helical (QH) equilibria \citep{paul2023fast} established that increased helicity of constant magnetic field strength $B_0$ lines in the QH case suppresses stochastic ion transport relative to QA stellarators. This study also demonstrated that, in stellarator FPPs, SAW perturbations resonant with passing fusion-born alpha particles can drive significant prompt loss of these particles. However, the study used ad hoc shear Alfv{\'e}n perturbations, whereas self-consistent transport assessment requires perturbations compatible with equilibrium plasma parameters. The minimal model for shear Alfv{\'e}n waves in stellarators is given by the ideal, incompressible vorticity model, implemented in the AE3D code \citep{spong2010clustered}. Furthermore, given that much recent research is focused on quasi-isodynamic (QI) designs \citep{goodman2024quasi}, which are closer to quasi-poloidal (QP) symmetry, it is instructive to extend the previous QA and QH resonance analysis \citep{paul2023fast} to the general quasi-symmetric (QS) case. In this work, we extend previous resonance analysis to general QS equilibria and investigate prompt stochastic ion losses driven by self-consistent shear Alfv{\'e}n perturbations.

The rest of the paper is organized as follows. In \S~\ref{sec:model_description}, we discuss the SAW perturbation model, the drift transport model, and the diagnostic tools \citep{duignan2023distinguishing} for analyzing particle stochasticity. In \S~\ref{sec:analysis}, we extend the previous analysis to the case of general quasisymmetry, and apply the presented assessment framework to several optimized stellarator equilibria. In \S~\ref{sec:conclusion}, we summarize and conclude.

\section{Model description}
\label{sec:model_description}

This section describes a drift kinetic, ideal reduced MHD model of ion transport in stellarators, in the presence of SAWs consistent with the equilibrium $B_0$ field and Alfv{\'e}n velocity $v_{\rm A}$ profile. In this section and throughout the paper, we use SI units for dimensional quantities.

The guiding center dynamics of a particle of mass $M_{\rm s}$ and charge $q_{\rm s}$, moving in the equilibrium magnetic field $\mathbf{B}_0=\nabla\times\mathbf{A}_0$ with vector potential $\mathbf{A}_0$  in the presence of a shear-Alfv{\'e}n wave perturbation with magnetic field $\delta\mathbf{B}=\nabla\times\alpha\mathbf{B}_0$, is described by the Lagrangian \citep{littlejohn1983variational}
\begin{align}
\label{eq:Littlejohn}
L(\mathbf{R},\frac{d{\mathbf{R}}}{dt},v_{\|},t)=q_{\rm s}(\mathbf{A}_0+\alpha \mathbf{B}_0+\frac{M_{\rm s}v_{\|}}{q_{\rm s}B_0}\mathbf{B}_0)\cdot\frac{d{\mathbf{R}}}{dt}-\frac{M_{\rm s}v_{\|}^2}{2}-\mu B_0-q_{\rm s}\delta \Phi,
\end{align}
where $\mathbf{R}$ is the position of the particle, $t$ is time, $B_0$ is the equilibrium magnetic field strength, $v_{\|}=\mathbf{v}\cdot\mathbf{B}_0/B_0$ is the component of the particle's velocity $\mathbf{v}$ along $\mathbf{B}_0$, $\mu=M_{\rm s}(v^2-v_{\|}^2)/(2{B}_0)$ is the magnetic moment, and $\delta \Phi$ is the scalar potential of the shear Alfv{\'e}n wave. The electric field of the transverse SAW is normal to $\mathbf{B}_0$, so the scalar $\delta \Phi$ and vector $\alpha \mathbf{B}_0$ potentials are related by the condition
\begin{align}
\label{eq:no_parallel_field}
\mathbf{B}_0\cdot\left(\frac{\partial\alpha\mathbf{B}_0}{\partial t}+\nabla\delta\Phi\right)=0\Rightarrow \nabla_{\|}\delta\Phi=-B_0\frac{\partial \alpha}{\partial t},
\end{align}
where $B_0\nabla_{\|}=\mathbf{B}_0\cdot\nabla$ is the derivative along the magnetic field. Because of~(\ref{eq:no_parallel_field}), the transverse SAW is fully described by its scalar potential $\delta \Phi$. In the flute-reduced $k_\|/k_\perp\ll 1$ incompressible ideal MHD model \citep{kruger1998generalized}, $\delta\Phi$ satisfies the Alfv{\'e}n vorticity equation 
\begin{align}
\label{eq:vorticity}
\omega^2 \nabla\cdot\left(\frac{1}{v_{\rm A}^2}\nabla\delta\Phi\right)+ B_0\nabla_{\|}\left[\frac{1}{ B_0}\nabla^2\left(\nabla_{\|}\delta\Phi\right)\right]=0,
\end{align}
where $k_\|$ and $k_\perp$ are wave vector components along and normal to the equilibrium field $\mathbf{B}_0$, $\omega$ is the frequency of the shear-Alfv{\'e}n wave, and $v_{\rm A}^2=B_0^2/(\mu_0 \rho_0)$, where $\rho_0$ is the plasma mass density and $\mu_0$ is the vacuum permeability. The vorticity eigenvalue problem~(\ref{eq:vorticity}) with the conducting shell $\delta\Phi=0$ boundary condition is the minimal model for SAW waves; it represents charge conservation $\nabla\cdot\mathbf{j}=0$ in ideal MHD, where $\mathbf{j}$ is the plasma current, and neglects compression $\gamma=0$ effects from the finite specific heats ratio $\gamma$.

Because the coupling between shear-Alfv{\'e}n and acoustic waves scales with plasma pressure \citep{cheng1985low}, neglect of compression effects in~(\ref{eq:vorticity}) is justified for the reduced MHD ordering model when the ratio of plasma pressure to the magnetic field pressure $\beta$ satisfies $\beta\lesssim k_\|/k_\perp$. Larger $\beta$ values can~\citep{turnbull1993global} lead to a pressure-induced gap in the continuum and cause mixing of SAW and acoustic branches.

The Ampere's law response in~(\ref{eq:vorticity}) is simplified by assuming that the wave vector's components along $k_\|$ and normal $k_\perp$ to the equilibrium $\mathbf{B}_0$ field are flute-ordered, $k_\|/k_\perp\ll 1$. Furthermore, the parallel equilibrium current $\mathbf{j}_0$ energy source is neglected to exclude kink instabilities, leaving only unconditionally stable shear Alfv{\'e}n waves with frequency $\omega$ given by,
\begin{align}
\omega^2=\frac{\int d^3\mathbf{r}|\nabla\nabla_\| \delta\Phi|^2}{\int d^3\mathbf{r}|\nabla\delta\Phi|^2/v_{\rm A}^2}>0.
\end{align}
It is convenient to solve the vorticity equation~(\ref{eq:vorticity}) in Boozer coordinates \citep{garren1991magnetic} for the equilibrium field $\mathbf{B}_0$, because the derivative $\nabla_\|$ along $\mathbf{B}_0$ has a particularly simple form in these coordinates. Therefore, ideal MHD (e.g., the AE3D code described in \citep{spong2010clustered} and gyrofluid (e.g., the FAR3D code described in \citet{varela2024stability}) codes represent the wave potential as a superposition of Fourier harmonics in Boozer angles,
\begin{align}
\label{eq:deltaPhi}
\delta\Phi=\sum_{m,n}\Phi_{m,n}(s)e^{i(m\theta-n\zeta+\omega t+\varphi_{m,n})},
\end{align}
where the sum is over harmonics with phase $\varphi_{m,n}$, radial profile $\Phi_{m,n}(s)$, and with poloidal $\theta$ and toroidal $\zeta$ Boozer angle mode numbers $m$ and $n$, respectively. The radial profile of the harmonic amplitude $\Phi_{m,n}(s)$ is a function of $s=\psi/\psi_{\rm total}$, the toroidal flux $\psi$ normalized by its total value $\psi_{\rm total}$. In Boozer coordinates, the co- and contravariant forms of the equilibrium $\mathbf{B}_0$ field are, respectively,
\begin{align}
\mathbf{B}_0=\psi_0\nabla s\times\nabla\theta-\iota\psi_0\nabla s\times\nabla\zeta=K\nabla s+I(s)\nabla\theta+G(s)\nabla \zeta,
\end{align}
where the rotation transform $\iota=d\psi_{\rm p}/d\psi$ is the rate of change of poloidal $\psi_{\rm p}$ with respect to toroidal $\psi$ flux. 
In what follows, the equilibrium current $\mathbf{j}_0=\nabla\times\mathbf{B}_0$ is assumed to be straight in Boozer coordinates. Consistent~\citep{imbert2024introduction} with the low-$\beta$ assumption used for neglecting compressibility effects, we neglect~\citep{white2017theory} the radial covariant $K=0$ component of the magnetic field. 

In stellarators, poloidal current $G$ is commonly \citep{helander2014theory} larger than the toroidal current $I$. Furthermore, $I\ll G$ follows from the near axis expansion~\citep{landreman2018direct} in stellarators with large aspect ratio $A=R/a\gg1$, where $a$ and $R$ are the minor and major stellarator radii, respectively. When $I\ll G$, the ratio of the harmonic's wave vector $\mathbf{k}=m\nabla\theta-n\nabla\zeta$ components that are parallel $k_\|$ and normal $k_\perp$ to the equilibrium field $\mathbf{B}_0$ can be expressed as 
\begin{align}
\label{eq:aspect_ratio}
\frac{k_{\|}^2}{k_{\perp}^2}\approx\frac{1}{A^2}\left(\iota-\frac{n}{m}\right)^2.
\end{align}
Analysis of the Alfv{\'e}n continuum~\citep{Paul_Hyder_Rodriguez_Jorge_Knyazev_2025} shows that centers of spectral gaps in the SAW spectrum have frequencies on the order of Alfv{\'e}n transit frequency $\omega / \omega_{\rm A}\sim1$, where $\omega_{\rm A}=v_{\rm A}/R=2\pi v_{\rm A} B_0^{\rm axis}/G_0^{\rm axis}$, and where $B_0^{\rm axis}$ and $G_0^{\rm axis}$ are the magnetic field strength and poloidal current on the magnetic axis. Since radially global SAWs are expected to reside in these spectral gaps, from the Alfv{\'e}n wave dispersion~\citep{Paul_Hyder_Rodriguez_Jorge_Knyazev_2025} relation ${|\iota m - n| = \omega / \omega_{\rm A}\sim1}$ it follows from~(\ref{eq:aspect_ratio}) that the flute-reduced~$k_\|/k_\perp\ll1$ vorticity equation~(\ref{eq:vorticity}) is valid for large aspect ratio $A\gg1$ equilibria.

\begin{figure}
    \centering
    \includegraphics[width=1.0\linewidth]{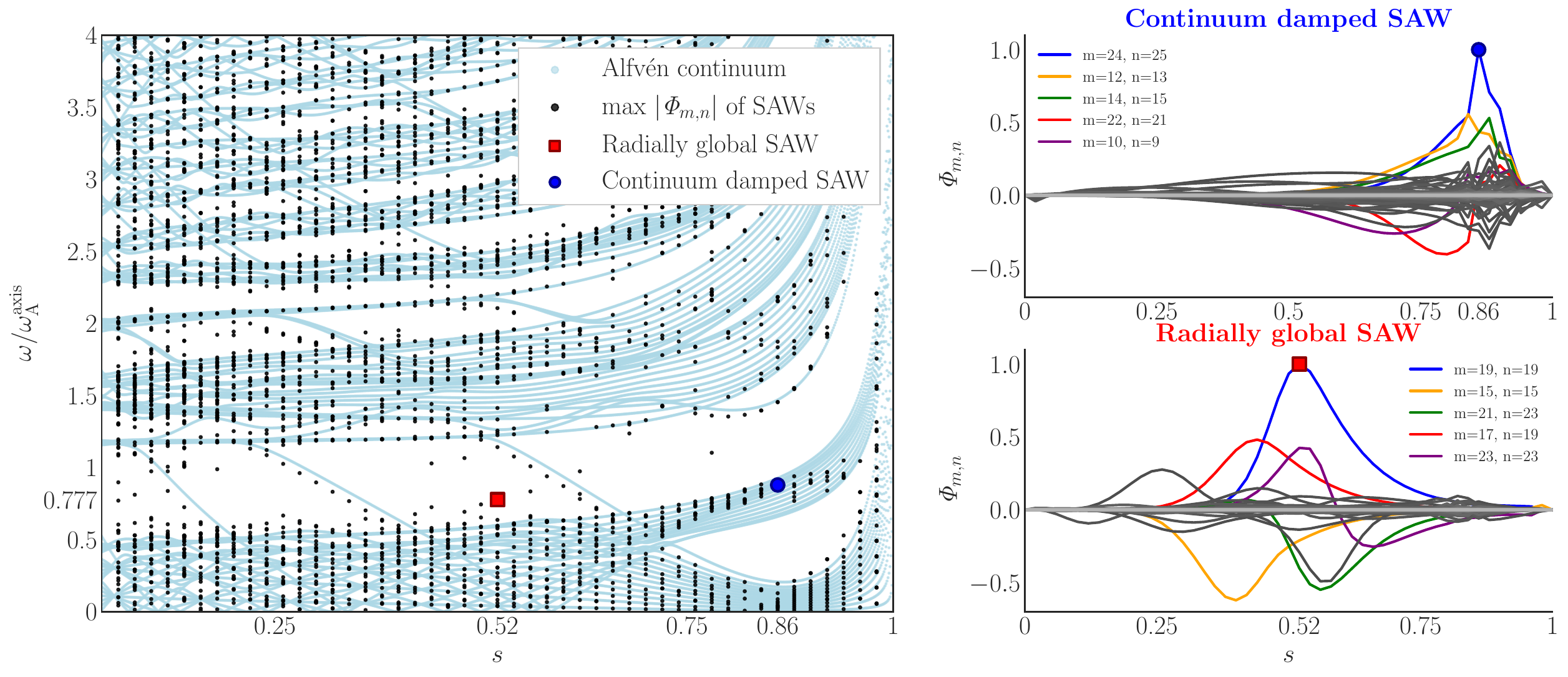}
    \caption{Example of AE3D wave classification based on STELLGAP continuum, for finite $\beta$ QH equilibrium \citep{landreman2022optimization} considered in \S~\ref{sec:analysis}. The left panel shows the Alfv{\'e}n continuum computed with STELLGAP, overlaid with AE3D results. Each black dot corresponds to a SAW from the AE3D calculation, placed at the radial location of the peak $|\Phi_{m,n}|$ amplitude across all harmonics~(\ref{eq:deltaPhi}). Most of these dots follow the continuum curves and identify continuum-damped SAW. The top-right panel shows one such SAW, with its peak at $s=0.86$ (blue circle) at the continuum crossing. Because the vorticity model~(\ref{eq:vorticity}) does not resolve continuum damping, the wave has abrupt radial structure on the scale of the AE3D radial grid. In contrast, the bottom-right panel shows a radially global SAW with frequency $\omega=34.3\;{\rm kHz}=0.777\omega_{\rm A}^{\rm axis}$, with its peak at $s=0.52$ (red square), in the gap of the Alfv{\'e}n continuum. In both right panels, colored lines show the five Fourier harmonics~(\ref{eq:deltaPhi}) with the largest peak $\Phi_{m,n}$ amplitude, with $m$ and $n$ mode numbers labeled in the legend. Gray lines show the remaining lower-amplitude harmonics from the simulation.}
    \label{fig:STELLGAP_AE3D}
\end{figure}

The ideal vorticity model~(\ref{eq:vorticity}) is fast to evaluate and only depends on the equilibrium field $\mathbf{B}_0$ and Alfv{\'e}n speed profile $v_{\rm A}$. However, it does not resolve the physics of continuum damping \citep{tataronis1973decay}. Because the ideal vorticity model~(\ref{eq:vorticity}) does not contain a physical scale for Alfv{\'e}n resonance, its numerical solutions can be subject to continuum damping and exhibit abrupt radial structure on the scale set by the grid spacing between the flux surfaces, as illustrated on the top right panel of figure~\ref{fig:STELLGAP_AE3D}. The frequencies of the resonant SAWs form a continuous spectrum, which can be modeled using the STELLGAP~\citep{spong2003shear} code. In gaps of this continuum, there exist discrete frequency modes not subject to continuum damping.

Solutions of~(\ref{eq:vorticity}) from the AE3D code can be conveniently classified as resonant or global by placing them over the corresponding continuum computed with STELLGAP. The result is illustrated in figure~\ref{fig:STELLGAP_AE3D}, where each AE3D solution is represented by its frequency $\omega$ at the location of the maximum $\Phi_{m,n}(s)$ for its most energetic harmonic.

Including resistivity resolves~\citep{timofeev1979theory} the continuum damping and results in decay of the resonant harmonics. Furthermore, accounting for inverse Landau damping interaction between the SAW and fast ion population~\citep{heidbrink2008basic} allows identification of Alfv{\'e}n modes destabilized by the energetic particle (EP) drive. The resistive and kinetic effects described here are included in gyrofluid code FAR3D~\citep{varela2024stability} and hybrid MHD-kinetic code M3DC-1K~\citep{liu2022hybrid}. Since the global SAW solutions of~(\ref{eq:vorticity}) are not subject to continuum damping, they are likely to be destabilized by the EP drive.

The equations of motion corresponding to the Lagrangian~(\ref{eq:Littlejohn}) can be solved in Boozer coordinates with the FIRM3D code~\citep{firm3d2025}, allowing modeling of the loss fraction induced by the wave perturbation~(\ref{eq:deltaPhi}). Since model~(\ref{eq:Littlejohn}) neglects collisional effects, it is most suited for evaluating prompt losses occurring faster than the collisional timescale. For fusion-born alpha particles in FPP reactors, the collisional timescale is set~\citep{helander2005collisional} by the slowing down frequency from the electron friction, $1/\nu_{\rm s}^{\alpha e}$. For FPP relevant plasmas similar to ARIES-CS studies \citep{bader2021modeling}, this slowing-down timescale is much larger than the Alfv{\'e}n transit time, $\omega_{\rm A}\gg \nu_{\rm s}^{\alpha e}$. Accordingly, alphas that promptly escape the plasma volume on the $1/\omega_{\rm A}$ timescale do not contribute to plasma heating and thus do not contribute to the collisional bulk heating necessary for sustained fusion burn.

We choose the fusion birth distribution function as uniform in velocity space. The density of the fusion-born alphas corresponds to the reactivity profile~\citep{bader2021modeling}
\begin{align}
\label{eq:bader_reactivity}
R_{\rm DT}&=n_{\rm D} n_{\rm T}\langle \sigma v \rangle = \frac{n_{\rm e}^2}{4}\langle\sigma v\rangle,
\end{align}
where $n_{\rm D}$, $n_{\rm T}$ and $n_{\rm e}$ are deuterium, tritium and electron density, and $\langle\sigma v\rangle$ is the reactivity cross section.  The cross-section we adopt for this study is from the recent~\citep{bader2021modeling} alpha particle transport assessment, and is given by
\begin{align}
\langle\sigma v \rangle &= \frac{3.6\times 10^{-18}\;{[{\rm keV}^{2/3}]}}{T_0^{2/3}}\exp\left(\frac{-19.94\;[{\rm keV}^{1/3}]}{T_0^{1/3}}\right)\;{\rm m^{3} sec^{-1}},
\end{align}
where the plasma temperature $T_0$ is in ${\rm keV}$.
\section{Fast ion motion and transport analysis}
\label{sec:analysis}

This section outlines three complementary techniques we used for the transport assessment. We first analyze resonances between passing fast ions and SAWs in the simplest case of a precisely quasi-symmetric field and a single harmonic perturbation in Boozer coordinates. For this configuration, we can determine the resonant locations and the onset of stochasticity analytically, and the results are discussed in \S~\ref{sec:stochastic_dynamics}. Stochasticity of particles in the described setup can be studied with the kinetic Poincar{\'e} cross-section technique similar to tokamaks, as discussed in \S~\ref{sec:KineticPoincare}. For general trajectories, their relative stochasticity can be compared with the weighted Birkhoff averaging technique outlined in \S~\ref{sec:WeightedBirkhoffAveraging}.

We apply these techniques to assess fast ion transport in several optimized stellarator equilibria. Similar to previous energetic particle studies \citep{bader2021modeling, paul2023fast}, we model losses of fusion-born alpha particles in equilibria scaled to ARIES-CS \citep{goodman2024quasi} minor radius $a=1.7\;{\rm m}$ and average magnetic field strength $B_0=5.7\;{\rm T}$ values. We use plasma profiles from \citep{bader2021modeling}, with a linear temperature profile $T_0=T_0^{\rm axis}(1-s)$, $T_0^{\rm axis}=11.5\;{\rm keV}$, and an equal-part deuterium-tritium mixture density on axis $n_0^{\rm axis}=2.25\times10^{20}\;{\rm m^{-3}}$ corresponding to a mass density $\rho_0^{\rm axis}=1.88\times10^{-6}\;{\rm kg/m^3}$, with radial density profile $\rho_0$ given in Table~\ref{tab:equilibria}.

For each equilibrium, the shear Alfv{\'e}n continuum and Alfv{\'e}n vorticity eigenproblems are calculated using STELLGAP~\citep{spong2003shear} and AE3D~\citep{spong2010clustered} codes.
The transport of the fusion-born alpha particle distribution, consistent with reactivity (\ref{eq:bader_reactivity}), is modeled using Monte Carlo simulations for various perturbation amplitudes, identifying the mode amplitude necessary to drive prompt fast ion losses.

For QA and QH equilibria, we consider $\beta=2.5\%$ equilibria from \citep{landreman2022optimization} as representative cases. For QI equilibrium, we consider a $\beta=2\%$ equilibrium SQuID similar to \citep{goodman2024quasi}. On each flux surface, the equilibrium $B_0$ field strength of these equilibria is represented as a superposition of quasisymmetric harmonics $B_{m,n}(m\theta-n\zeta)$, summed over integers $m$ and $n$. Although none of the equilibria are exactly QS, each has a dominant $B_{m,n}$ component across the entire volume. The $M$ and $N$ values of the dominant $B_{m,n}$ for each equilibrium are summarized in Table~\ref{tab:equilibria}.

\begin{table}
\centering
\caption{Summary of stellarator equilibria parameters scaled to ARIES-CS size.}
\label{tab:equilibria}
\begin{tabular}{lccccccc}
\hline
Class & $\beta$ (\%) & $N_{\rm fp}$ & \makecell{$m$, $n$ for \\ dominant $B_{m,n}$}  & $B_0^{\rm axis}$ [T] & $\iota_{\rm min}$, $\iota_{\rm max}$ & \makecell{Aspect ratio\\  $A=R/a$} & \makecell{$\rho_0/\rho_{0}^{\rm axis}$ \\profile}\\
QA   &  2.5  & 2 &  1, 0  & 5.9 & $0.19$, $0.48$ &6.0 & $1$\\
QH   &  2.5  & 4 &  1, 4  & 5.9 & $1.01$, $1.16$ &6.5 & $1-s^5$\\
QI   &  2    & 4 &  0, 4  & 6.0 & $0.70$, $0.98$ &10.0 &$1-s^5$\\
\end{tabular}
\end{table}
\begin{figure}
    \centering
    \includegraphics[width=1.0\linewidth]{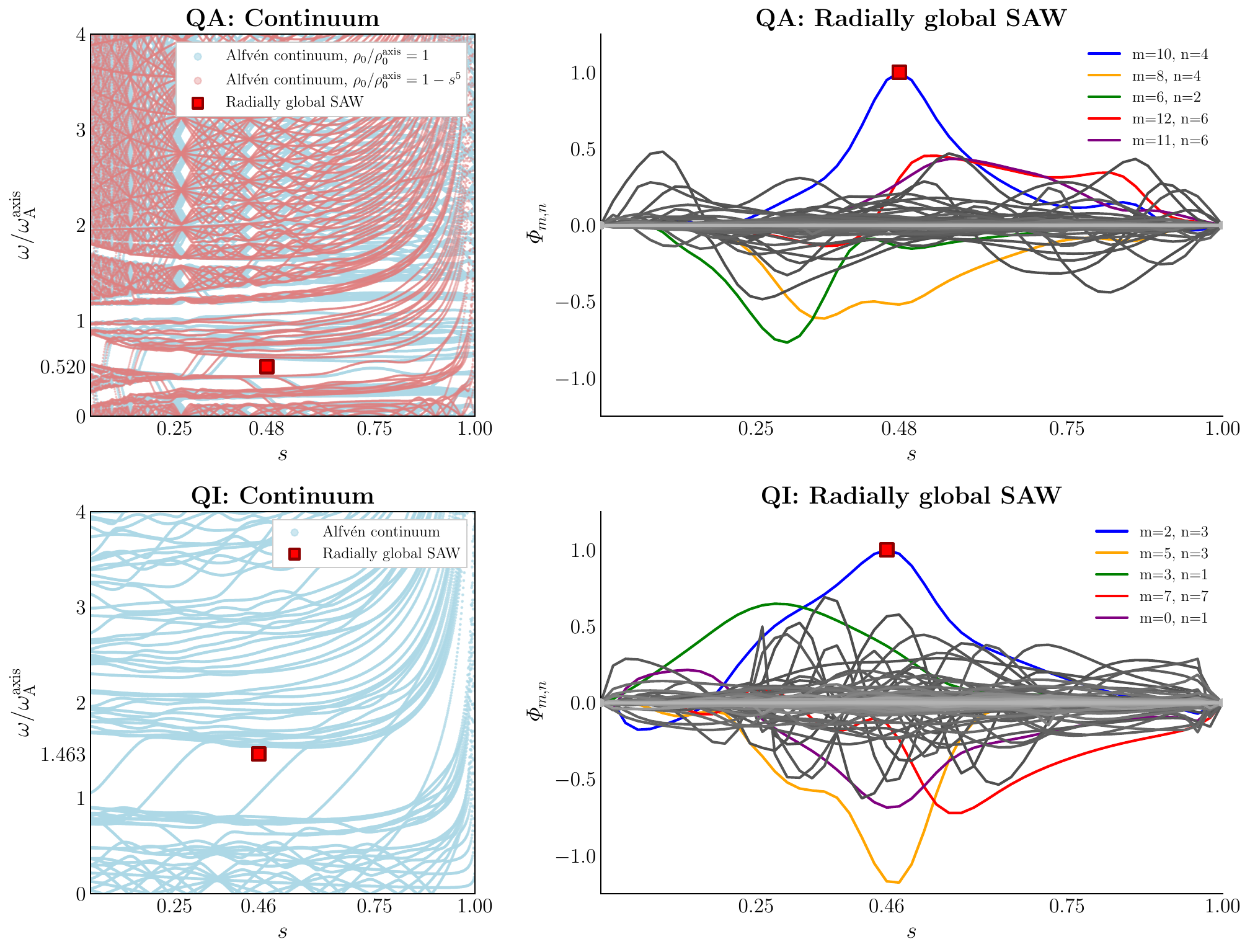}
    \caption{Alfv{\'e}n continuum (left) and radially global SAW eigenfunctions (right) for QA (top, 30.3 kHz) and QI (bottom, 46.8 kHz) configurations from Table~\ref{tab:equilibria}, computed with STELLGAP and AE3D. Black dots indicate mode amplitude maxima; red squares mark the highlighted global modes. As in figure~\ref{fig:STELLGAP_AE3D}, colored lines show the five Fourier harmonics~(\ref{eq:deltaPhi}) with the largest peak $\Phi_{m,n}$ amplitude, with $m$ and $n$ mode numbers labeled in the legend. Gray lines show the remaining lower-amplitude harmonics from the simulation. The narrow gaps in QA lead to strong continuum damping in the presence of the density shear, whereas the wide helical gaps in QI support radially extended global SAW that persist despite the edge continuum shear near $s \sim 1$ caused by the $\rho_0^{\rm axis}(1-s^5)$ density profile.}
    \label{fig:STELLGAP_AE3D_QA_QI}
\end{figure}
The STELLGAP Alfv{\'e}n continua calculations, and the global SAW modes from AE3D for equilibria in Table~\ref{tab:equilibria} are shown in figure~\ref{fig:STELLGAP_AE3D} for QH, and in figure~\ref{fig:STELLGAP_AE3D_QA_QI} for QA and QI stellarators. Dynamic mode resolution parameters for all simulations are outlined in Appendix~(\ref{sec:simulations}). Note that the sheared $\rho_0^{\rm axis}(1-s^5)$ density profile used in QH and QI equilibria results in steepening near the edge $s\sim 1$, characteristic of the Alfv{\'e}n velocity profile shear $v_{A}^2\sim B_0^2/\rho_0$ due to a decrease in plasma density $\rho_0$ near the last flux surfaces. Accordingly, the radially global SAWs residing in the continuum frequency gaps have most of their energy localized away from the device edge. 

Due to much narrower frequency gaps in the QA continuum compared to QH and QI continua shown in figures~\ref{fig:STELLGAP_AE3D}~and~\ref{fig:STELLGAP_AE3D_QA_QI}, the QA equilibrium modes are more sensitive to continuum damping induced by the density shear. For the QI and the QH equilibria, AE3D finds global modes that persist in both flat and sheared $\rho_0^{\rm axis}(1-s^5)$ plasma density profiles, with radial structure unaffected by the difference in Alfv{\'e}n velocity shear in these two cases. For the QA equilibria, none of the radially global modes found in flat profiles persisted in sheared $\rho_0^{\rm axis}(1-s^5)$ density profile because of the continuum damping near the edge. This supports continuum gap width optimization \citep{Paul_Hyder_Rodriguez_Jorge_Knyazev_2025} as a viable strategy for SAW suppression. In the case of a flat density profile, radial Alfv{\'e}n velocity shear comes from the variation in the magnetic field strength $B_0$, and the overall shear is decreased relative to the $\rho_0^{\rm axis}(1-s^5)$ profile. We find a SAW mode in the QA configuration with a flat density profile, and use it in our analysis.

The rest of the section will use the equilibria and global SAWs described above to illustrate and verify the analysis. We characterise the magnitude of SAW perturbation by the magnetic field energy ratio between the wave $\delta U$ and the equilibrium $U_0$,
\begin{align}
\label{eq:saw_amplitude}
\frac{\delta U}{U_0}=\frac{\int d\mathbf{x} \delta B^2}{\int d\mathbf{x}  B_0^2}.
\end{align}
As energy is quadratic in wave amplitude, $\sqrt{\delta U/U_0}$ serves as a measure for average magnetic field perturbation in stellarator volume. Another common \citep{nazikian1997alpha, paul2023fast} way to characterize SAW amplitude is the normalized maximum of the magnetic field perturbation,
\begin{align}
\label{eq:normalized_Bs}
\delta \hat B^s=\frac{\delta \mathbf{B}\cdot\nabla s}{B_0|\nabla s|}.
\end{align}
For SAWs and equilibria we consider, prompt losses for different wave amplitudes $\sqrt{\delta U/U_0}$ are shown in figure~\ref{fig:harmonics_and_losses}. Measured prompt EP losses exceed 1\% for medium $\sqrt{\delta U/U_0}$ (maximum radial~(\ref{eq:normalized_Bs})) magnetic field of $1\times10^{-3}$ ($3.1\times10^{-3}$) in QA, $2.15\times10^{-3}$ ($6.6\times10^{-3}$) in QH, and $2.15\times10^{-3}$ ($3.3\times10^{-3}$) in QI. This is consistent with the $2\times10^{-3}$ threshold reported for TFTR modeling~\citep{hsu1992alpha}. Notably, threshold magnetic fields across DT tokamak analyses vary significantly, with TFTR experiments~\citep{nazikian1997alpha} reporting prompt losses with inferred normalized field amplitudes of $\delta\hat B^s\sim10^{-5}$, while recent ITER prompt loss analysis~\citep{van2012alfven} suggests negligible losses for amplitudes up to $\delta\hat B^s\sim10^{-3}$. Experiments from LHD~\citep{ogawa2012magnetic} observe prompt losses for similar $\delta\hat B^s\sim10^{-4}-10^{-3}$ inferred SAW amplitudes.

\subsection{Passing resonance analysis for single harmonic SAW}

\label{sec:stochastic_dynamics}
To better understand the onset of stochastic transport caused by the presence of the shear Alfv{\'e}n wave, it is instructive to consider the simplest case of a precisely quasi-symmetric equilibrium with a perturbation mode consisting of a single Fourier harmonic $(m,n)$ in Boozer coordinates. Under these assumptions, the dynamics of a passing particle in a stellarator can be analyzed similarly to the tokamak case \citep{mynick1993stochastic}, as was done \citep{paul2023fast} to show that the field-strength helicity of QH equilibria suppresses stochastic transport. In what follows, we generalize the analysis done in \citep{paul2023fast} to the case of arbitrary helicity symmetry. The extended analysis includes the quasi-poloidal case, which is appropriate for approximating QI equilibria.

The resonant analysis below relies on several assumptions regarding equilibrium unperturbed particle dynamics and the perturbation field:

\begin{assumptions}
\item \label{assump:quasisym} \textbf{Quasi-symmetric magnetic field:} The equilibrium magnetic field is assumed to be quasi-symmetric, i.e.\ $B_0(s,\theta,\zeta)=B_0(s,\chi=M\theta-N\zeta)$, where $M$ and $N$ are integers specifying the field helicity. Current techniques of stellarator optimization~\citep{landreman2022magnetic} achieve sufficient quasi-symmetry precision to justify this assumption. This analysis can also be useful for QI equilibria that can be approximated by QP symmetry.

\item \label{assump:singlepert} \textbf{Single harmonic perturbation:} Similar to the earlier works~\citep{mynick1993stochastic,paul2023fast}, the perturbation is taken as a single Fourier harmonic, which provides continuous symmetry in the problem and a resulting motion invariant. This motion invariance is absent in more realistic mode structures consisting of several Fourier harmonics in Boozer coordinates, which significantly complicates the analysis \citep{white2012modification}.

\item \label{assump:passing} \textbf{Passing particle assumption:} For passing particles, we can express the unperturbed motion as a Fourier series in the symmetry angle $\chi$ for the resonance analysis (see Appendix~\ref{sec:appendix} for details). The same approach does not work for trapped particles that have consecutive bounces $\Delta\chi_{\rm b}$ in the same field period $\Delta\chi_{\rm b}<2\pi$. However, our conclusions will still hold for barely trapped particles $\Delta\chi_{\rm b}\gg2\pi$. Notably, passing particles also have a one-to-one correspondence between the particle's energy $E$ and its parallel velocity $v_{\|}$. This enables one to construct Poincar{\'e} cross-sections of perturbed EP motion, discussed in \S~\ref{sec:poincare_crossection}, to verify our resonance analysis, as shown in figure~\ref{fig:Poincare_QIh_WBA}.

\item \label{assump:singleharm} \textbf{Single Fourier harmonic in unperturbed trajectory:} For passing particles (assumption~\ref{assump:passing}), we further assume that the Fourier expansion of the trajectory in the symmetry angle $\chi$ has a single dominant harmonic. This is consistent with QA and QH equilibria: the near-axis~\citep{landreman2018direct} expansion,
\begin{align}
\label{eq:nae}
B_0(s,\chi)= B_0^{\rm axis}(1-\bar\xi\sqrt{s} \cos(\chi)),\quad\bar\xi=const,
\end{align}
guarantees a single dominant field strength Fourier harmonic near the axis, and the corresponding $\nabla B_0$ drift then produces a dominant Fourier harmonic in the particle's trajectory. Although the same near-axis expansion~(\ref{eq:nae}) does not apply to QI equilibria, they can still have a dominant harmonic in Boozer components. One such example is the QI equilibrium considered in Table~\ref{tab:equilibria}, where a single poloidal $M=0$ harmonic dominates equilibrium field $B_0$ across the stellarator volume, as illustrated in figure~\ref{fig:harmonics_and_losses}.

\item \label{assump:vacuum} Lastly, for simplicity, we consider a \textbf{vacuum} $I=0$ field. This is consistent with the low-$\beta$ assumption and the near-axis expansion of stellarator equilibrium.

\end{assumptions}
Under the listed (\ref{assump:quasisym}--\ref{assump:vacuum}) assumptions, it can be shown (see Appendix~\ref{sec:appendix}) that the $(m,n)$ SAW harmonic of frequency $\omega$ resonates with the unperturbed drift orbit when
\begin{align}
\label{eq:resonant_condition}
\Omega_{l}&= \omega+(m+lM)\omega_\theta-(n+lN)\omega_\zeta=0.
\end{align}
This resonance condition (\ref{eq:resonant_condition}) can be expressed in terms of orbit helicity $h=\omega_\theta/\omega_\zeta$ as
\begin{align}
\label{eq:resonant_helicity}
h_l = \frac{n+lN-\omega/\omega_\zeta}{m+lM},
\end{align}
where $l$ is the integer describing the drift harmonic, $h_l$ is the corresponding resonant helicity, and $\omega_\theta$ ($\omega_\zeta$) is the time-averaged drift in poloidal (toroidal) Boozer angle,
\begin{align*}
\omega_\theta=\lim_{T\to\infty}\int_0^T\frac{\dot\theta dt}{T}, \quad \omega_\zeta=\lim_{T\to\infty}\int_0^T\frac{\dot\zeta dt}{T}.
\end{align*}
Neglecting~\citep{paul2023fast} radial dependence of average toroidal drift, $d\omega_\zeta/ds=0$, the island width at the resonant~(\ref{eq:resonant_helicity}) helicity $h_l$ has radial width
\begin{align}
\label{eq:resonant_width}
w_l^s\approx\sqrt{\frac{s_l}{(m+lM)d\omega_\theta/ds}},
\end{align}
where $s_l$ is the radial drift amplitude~(see (\ref{eq:drift_amplitudes}) in Appendix~(\ref{sec:appendix})).
It follows from~(\ref{eq:resonant_helicity}) that the spacing between the neighboring resonance helicities $h_l$ and ${h_{l+1}=h_l-\Delta s_l (dh_l/ds)}$ is
\begin{align}
\label{eq:island_width}
\Delta s_l=\frac{1}{dh_l/ds}\frac{Mh_0-N}{m+(l+1)M}.
\end{align}
Comparing the width $w_l^s$ of resonance islands~(\ref{eq:resonant_width}) to the spacing $\Delta s_l$ between resonances~(\ref{eq:island_width}) gives the island overlap condition~\citep{chirikov1971research} for the stochasticity onset,
\begin{align}
\label{eq:stochasticity_threshold}
\frac{w_l^s}{\Delta s_l}=\frac{m+(l+1)M}{(Mh_0-N)\omega_\zeta}
\sqrt{\left|\frac{s_l}{m+lM}\frac{d\omega_\theta}{ds}\right|}\gtrsim 1.
\end{align}
For a quasi-helical equilibrium, $\chi_{\rm QH}=\theta-N\zeta$, the stochasticity threshold criterion~(\ref{eq:stochasticity_threshold}) recovers results obtained in \citep{paul2023fast}. For a “quasi-poloidal” (QP) case, $\chi_{\rm QP}=-N\zeta$, the resonance condition~(\ref{eq:resonant_condition}) is
\begin{align}
\label{eq:poloidal_resonance}
\Omega_{l,{\rm QP}}=\omega+m\omega_\theta-(n+lN)\omega_\zeta=0,
\end{align}
which corresponds~(\ref{eq:resonant_helicity}) to helicity
\begin{align}
\label{eq:QP_resonant_helicity}
h_{\rm QP}=\frac{n+lN-\omega/\omega_\zeta}{m}.
\end{align}
The corresponding stochasticity threshold~(\ref{eq:stochasticity_threshold}) for QP is
\begin{align}
\label{eq:island_overlap}
\frac{w_{l,{\rm QP}}^s}{\Delta s_{l,{\rm QP}}}=-\frac{m}{N\omega_\zeta}\sqrt{\left|\frac{s_l}{m}\frac{d\omega_\theta}{d s}\right|}\gtrsim 1.
\end{align}
As in the QH case, stochasticity is suppressed for large $N$.  This corresponds to a poloidally symmetric field with a larger number of field periods $N_{\rm fp}$. Notably, QP has wider spacing between resonant islands~(\ref{eq:island_width}) than QH with the same sign of $Mh_0$ and $N$, leading~(\ref{eq:stochasticity_threshold}) to stochasticity suppression. The resonance (\ref{eq:poloidal_resonance}) with $m=0$ modes no longer depends on $\omega_\theta$: it occurs when $\omega/\omega_\zeta=n-lN$. In this case, the stochasticity onset is determined not by poloidal $d\omega_\theta/ds$ as in~(\ref{eq:island_overlap}) but rather toroidal $d\omega_\zeta/ds$ shear.


\subsubsection{Orbit classification and passing-trapped transitions}
\label{sec:orbit_classification}
\begin{figure}
\centering
\includegraphics[width=0.9\linewidth]{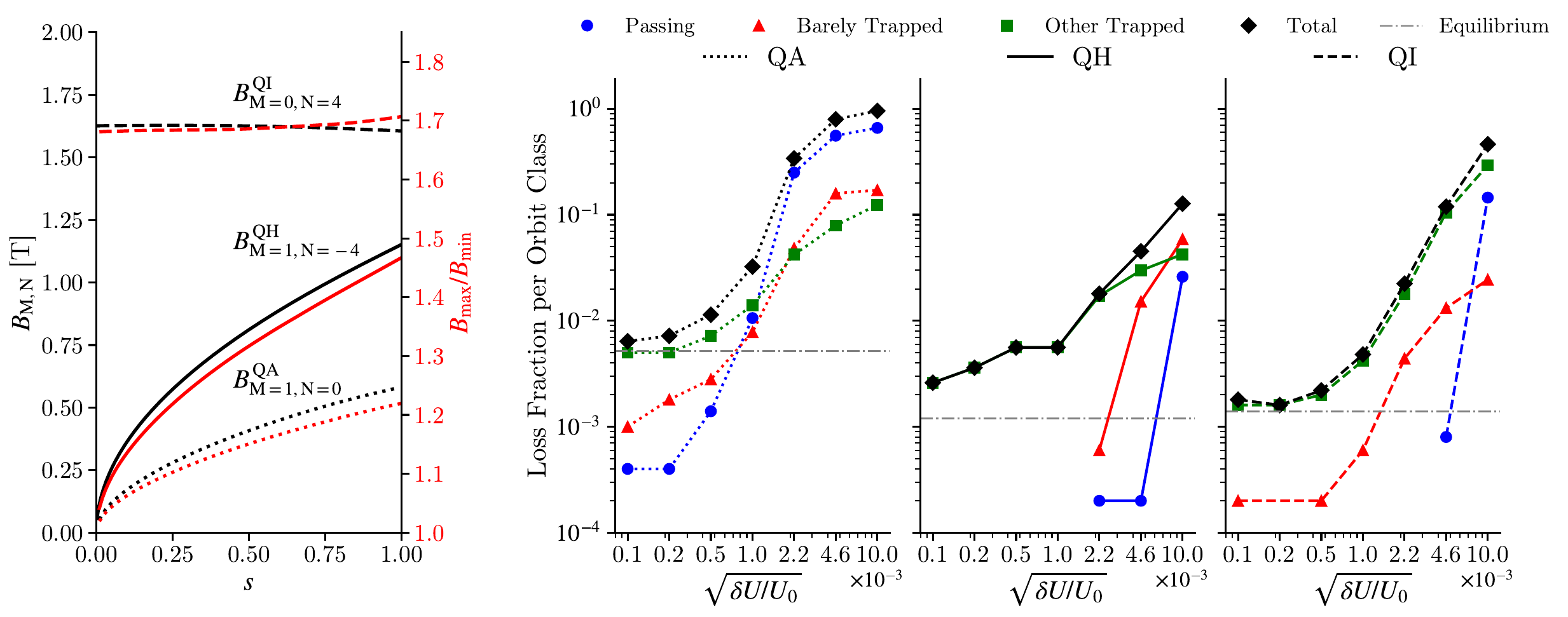}
\caption{The figure compares equilibrium fields and losses for QA (dotted), QH (solid), and QI (dashed) stellarators from Table~\ref{tab:equilibria}. Left panel shows (black) dominant Boozer harmonics $B_{M,N}$ and (red) mirror ratio $B_{\max}/B_{\min}$. The right three panels show prompt losses in QA, QH, and QI equilibria, for passing (blue circles), trapped (green squares), and barely-trapped (red triangles) orbit classes. Black diamond lines show total losses across orbit classes. Gray dash-dotted lines show losses without SAW.}
\label{fig:harmonics_and_losses}
\end{figure}

When the particle mirrors in a precisely QS field, consecutive bounces remain within the same period of the field symmetry angle, $\Delta \chi_{\rm b} < 2\pi$. However, SAW perturbations cause radial displacement away from the initial flux surface. A different peak magnetic field strength $B_0^{\rm max}$ at that new location means that a particle that is passing in the unperturbed field can mirror in the presence of SAW even after several periods, $\Delta \chi_{\rm b} > 2\pi$. This wave-induced transition between passing and barely trapped orbit classes can lead to prompt losses \citep{hsu1992alpha}. This loss channel is more pronounced for equilibria with larger radial variation in the mirror ratio. Notably, while it follows from the near-axis QS expansion (\ref{eq:nae}) that the QA and QH equilibria always have significant radial variation in their dominant $B$ harmonic, the radial variation of the magnetic field strength in QI is typically~\citep{velasco2023robust} small. This difference in radial $B_0$ variation is illustrated in figure~\ref{fig:harmonics_and_losses} and suggests that the transition between passing and trapped orbit classes is more pronounced in QA and QH than in QI. We test this numerically by considering \citep{paul2022energetic} the change in the symmetry angle $\chi$ between bounces of each lost particle: the passing particle never bounces, the ``barely trapped'' particles travel over a $2\pi$ period in $\chi$ between bouncing, and the remaining losses are particles trapped in the main harmonic of the field strength. For QI, we use the QP symmetry angle $\chi=-N\zeta$ for orbit classification. 
Consistent with this analysis, figure~\ref{fig:harmonics_and_losses} shows that in QS equilibria, significant SAW-induced losses occur for ``barely-trapped'' orbits. This is similar to the SAW-induced orbit transition loss mechanism in tokamaks~\citep{hsu1992alpha}. Due to the lower radial variance in magnetic field strength, this mechanism is suppressed in QI. Consistent with the resonant analysis~(\ref{eq:stochasticity_threshold}), passing particle losses are suppressed in QH and QI because field periods $N_{\rm fp}$ in these cases increase helicity $N$. Notably, the majority of losses in QI come from trapped particles. This motivates future work on extending resonant analysis beyond passing particles.
\subsection{Stochastic motion identification}
For identifying island overlap and stochasticity onset under assumptions of \S~\ref{sec:stochastic_dynamics}, we can use Poincar{\'e} cross-section technique described in \S~\ref{sec:poincare_crossection}. However, the assumptions (\ref{assump:quasisym}--\ref{assump:passing}) required to construct Poincar{\'e} cross-sections are too restrictive for comprehensive transport assessment. For a general trajectory, we measure stochasticity using weighted Birkhoff averaging technique described in \S~\ref{sec:WeightedBirkhoffAveraging}.

\subsubsection{Kinetic Poincar{\'e} cross-sections}
\label{sec:poincare_crossection}
Under assumptions (\ref{assump:quasisym}--\ref{assump:passing}) from the resonance analysis in \S~\ref{sec:stochastic_dynamics}, it is possible to construct~\citep{paul2023fast} a kinetic Poincar{\'e} cross-sections for perturbed particle motion. 

\begin{figure}
\centering
\includegraphics[width=0.9\linewidth]{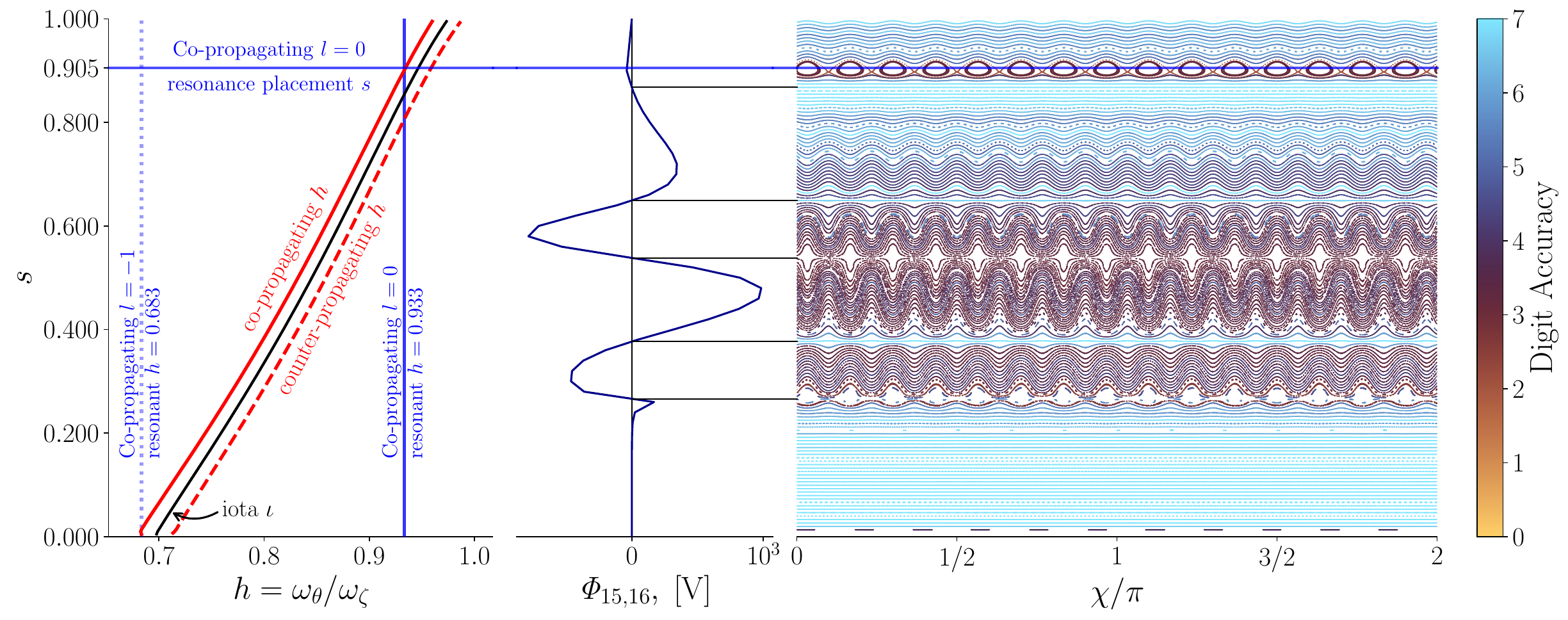}
\caption{The figure shows interaction between a single harmonic SAW and passing $\mu=0$ alpha particles co-propagating along the quasi-poloidal approximation to the QI field. The QI equilibrium is described in Table~\ref{tab:equilibria}. The perturbation is the most energetic harmonic of the $\omega=46.8\;{\rm kHz}$ SAW in the QI equilibrium, with poloidal $m=15$ and toroidal $n=16$ mode numbers. Left panel demonstrates resonance analysis (\ref{eq:poloidal_resonance}), showing that the mode resonates with the co-propagating trajectory of helicity $h=0.933$ near $s=0.905$ flux surface. This agrees with the Poincar{\'e} cross-section shown on the right panel, where there is an island chain at $s=0.905$. The middle panel shows the radial profile $\Phi_{15,16}(s)$ of the mode, with zero crossings $\Phi_{15,16}=0$ creating ``island-like'' structure on the Poincar{\'e} map. Because of the wide spacing between $l=0$ and $l=1$ resonances, the island overlap is avoided for this perturbation.}
\label{fig:Poincare_QIh_WBA}
\end{figure}

For constructing the Poincar{\'e} cross-section, note that the Boozer angle dependence of the SAW phase can be expressed in terms of the symmetry angle $\chi$ and some other angle $\eta=M_\eta\theta+N_\eta\zeta$, with $M_\eta$ and $N_\eta$ such that $s,\chi,\eta$ form a valid coordinate system. For precisely QS field, Lagrangian~(\ref{eq:Littlejohn}) becomes a function of $L(s,\chi,n_\eta\eta-\omega t,v_\|)$, where ${n_\eta=(mN-nM)/(MN_\eta-M_\eta N)}$. Therefore,
\begin{align}
\label{eq:EOM}
n_\eta\frac{\partial L}{\partial t}+\omega\frac{\partial L}{\partial \eta} = \frac{d}{dt}(n_\eta E-\omega P_\eta)=0,
\end{align}
where 
\begin{align}
E=\frac{M_{\rm s}v_\|^2}{2} +\mu B_0 + q_{\rm s}\delta \Phi
\end{align}
is the energy and 
\begin{align}
P_\eta=\frac{\partial L}{\partial \dot\eta}=\frac{q_{\rm s}(M\psi_{\rm p}-N \psi)-(NI+MG)\alpha}{MN_\eta-M_\eta N}-\frac{(NI+MG) M_{\rm s} v_\|}{(MN_\eta-M_\eta N)B_0}
\end{align}
is the canonical momentum conjugate to $\eta$. For fixed $n_\eta\eta-\omega t=\bar\eta=\mathrm{const}$, integral of motion~(\ref{eq:EOM}) gives a quadratic equation for $v_\|$ in terms of $\mu$, $s$, $\chi$ and $\bar\eta$. Since passing particles do not bounce, they preserve the sign of $v_\|$, so $v_\|$ can be determined (\ref{eq:EOM}) from values of $\mu, E', \chi$ and $s$. This allows one to construct Poincar{\'e} cross-section for passing particles of the same $\mu$, $E'$ values by recording their $s$ and $\chi$ values at fixed $\bar\eta$, as shown in figure~\ref{fig:Poincare_QIh_WBA}.

Poincar{\'e} cross-sections allow studying the resonances described by~(\ref{eq:resonant_condition}) and illuminate the impact of radial mode amplitude dependence $\Phi_{m,n}(s)$ on fast ion drift surfaces. An example in figure~\ref{fig:Poincare_QIh_WBA} shows drift surface deformation caused by the interaction with the mode. First is the resonance with the mode~(\ref{eq:resonant_condition}) described in the previous section. The second mechanism, also producing the island-like structures on the Poincar{\'e} map, is associated with the radial sign reversal of the wave’s electric $\delta \mathbf{E}$ and magnetic $\mathbf{\delta B}$ field. Recall that the wave’s contribution to EP radial $\delta s$ dynamics (\ref{eq:Littlejohn}) is because of (a) electric field of the mode causes radial $\mathbf{\delta E}\times\mathbf{B}_0$ drift and (b) the radial component of $\delta B$ displaces field lines from equilibrium flux surfaces $s=\mathrm{const}$, $\mathbf{B}\cdot\nabla s = \mathbf{\delta B}\cdot \nabla s$, causing particles streaming along field lines to drift radially. For a single harmonic SAW, these two effects are given by
\begin{align}
\delta \dot s=-\left(1+(\iota m -n)\frac{v_{\|}B_0}{\omega G}\right)\frac{m}{q_{\rm s}\psi_0}\Phi_{m,n}(s)\sin(m_\chi \chi - \bar\eta),
\end{align}
where $m_\chi=(mN_\eta-nM_\eta)/(MN_\eta-M_\eta N)$.
Near the location where $\Phi_{m,n}(s)=0$, the $\mathbf{E}\times\mathbf{B}$ drift from the wave acts in opposite directions. Similarly $\delta\mathbf{B}$ is also oppositely directed. The result is the island-like deformation of the drift surfaces near radial locations of zero wave amplitude. Figure~\ref{fig:Poincare_QIh_WBA} also shows that “island-like” chains for neighboring $\Phi_{m,n}(s)=0$ zeros are shifted in by $m_\chi\chi\sim\pi$ relative to each other, corresponding to changes in radial $d\Phi_{m,n}/ds$ slope.
\label{sec:KineticPoincare}

\subsubsection{Weighted Birkhoff Averaging}
\label{sec:WeightedBirkhoffAveraging}
\begin{figure}
\centering
\includegraphics[width=0.9\linewidth]{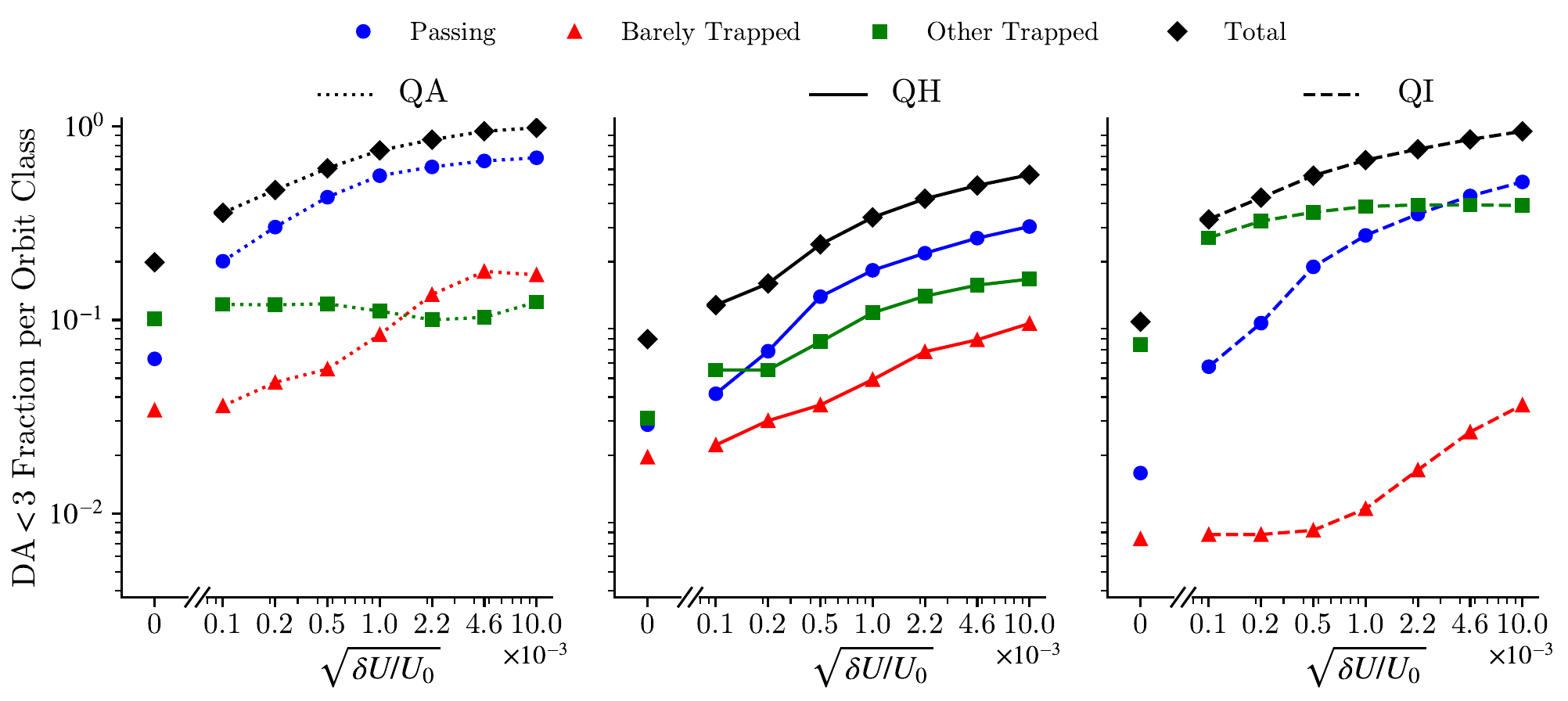}
\caption{Three panels show fraction of low digit accuracy ${\rm DA}<{\rm DA_{threshold}}=3$ fusion-born alpha particles in QA (left), QH (center), and QI (right) equilibria, for passing (blue circles), trapped (green squares), and barely-trapped (red triangles) orbit classes. Black diamond lines show total fraction across orbit classes.}
\label{fig:WBA_fraction}
\end{figure}
%
Another useful technique for assessment of particle's motion stochasticity is the recently proposed \citep{duignan2023distinguishing} weighted Birkhoff averaging (WBA) method. Unlike the Poincar{\'e} cross-section technique, the WBA approach does not rely on any of the (\ref{assump:quasisym}--\ref{assump:passing}) assumptions, making it suitable for analyzing SAW-induced stochasticity for all particles in non-QS equilibria with general Alfv{\'e}n wave perturbations.

The WBA implementation we use, detailed in~\citep{lachmann2025poincare}, classifies trajectories as integrable or chaotic based on the rate of convergence of the Birkhoff average of the particle's canonical momenta along its trajectory. This rate of convergence, quantified as the {\it digit accuracy} ${\rm DA}$, is higher for regular motion than for stochastic. Accordingly, the motion is classified as stochastic when its ${\rm DA}$ is below the threshold ${\rm DA}={\rm DA_{threshold}}$ value. For finding a suitable threshold ${\rm DA_{threshold}}$ value, it is instructive to evaluate ${\rm DA}$ on Poincar{\'e} cross-section, as illustrated in figure~\ref{fig:Poincare_QIh_WBA}. 
Although the computed value of DA can depend on implementation details~\citep{ruth2024finding}, such as the filtering function or numerical scheme used for modeling particle dynamics, analysis of FIRM3D tracing results~\citep{lachmann2025poincare} shows ${\rm DA_{threshold}}=3$ is a suitable classification threshold across configurations and perturbation amplitudes we consider. 

For assessment of shear Alfv{\'e}n wave-induced stochasticity of different orbit classes, it is instructive to compute their fraction of low ${\rm DA}<{\rm DA}_{\rm threshold}$ particles~\citep{lachmann2025poincare}. For each stellarator from Table~\ref{tab:equilibria}, we trace 5000 fusion-born alpha particles, classify them by orbit type and compute their digit accuracy. Resulting stochastic fractions for different SAW amplitudes are shown in figure~\ref{fig:WBA_fraction}.

In the absence of SAW perturbations, the WBA analysis in figure~\ref{fig:WBA_fraction} shows that most stochastic particles are trapped. This is consistent with recent analysis of phase-space integrability in quasisymmetric stellarators~\citep{chambliss2025fast}, which showed that trapped drift orbits are more susceptible to resonances with symmetry-breaking equilibrium perturbations than passing orbits. Notably, for sufficiently large SAW perturbations, the fraction of stochastic passing orbits exceeds that of trapped orbits. This suggests that, in contrast to the equilibrium perturbations considered in~\citep{chambliss2025fast}, time-dependent SAW perturbations primarily resonate with passing orbits, consistent with the resonance analysis of~\S~\ref{sec:stochastic_dynamics}.

For passing orbits, the stochastic fraction in the four-field-period $(N_{\rm fp}=4)$ QH and QI cases is lower than in QA. This agrees with the resonance analysis of~\S~\ref{sec:stochastic_dynamics}, where the increase in field helicity $N$
is shown to suppress island overlap~(\ref{eq:stochasticity_threshold}) in QH and QI but not QA. However, the island overlap condition does not account for the radial mode structure or the differences in dominant harmonic composition~(\ref{eq:deltaPhi}) between SAW in different equilibria. As shown in figures~\ref{fig:STELLGAP_AE3D} and~\ref{fig:STELLGAP_AE3D_QA_QI}, SAW modes can differ significantly across stellarators, complicating direct stochasticity comparisons. We note, for example, that while the resonance analysis predicts wider island spacing~(\ref{eq:island_width}) in the QI case than in QH for the same perturbation harmonic, figure~\ref{fig:WBA_fraction} shows no reduction in the stochastic passing orbit fraction for QI relative to QH.

Since stochasticity only leads to prompt losses when the stochastic region intersects the last closed flux surface, it is possible to have a large stochastic fraction with small prompt losses. For example, figure~\ref{fig:WBA_fraction} shows a large fraction of stochastic passing particles in the QH case for perturbations with average amplitude $\sqrt{\delta U/U_0}\gtrsim 10^{-4}$, yet figure~\ref{fig:harmonics_and_losses} shows that these particles do not contribute to prompt losses until the larger amplitude $\sqrt{\delta U/U_0}\gtrsim 10^{-3}$. This is especially relevant for SAW perturbations localized away from the edge due to radial Alfv{\'e}n velocity shear, as discussed in~\S~\ref{sec:analysis}.

Consistent with the orbit classification analysis of~\S~\ref{sec:orbit_classification}, figure~\ref{fig:WBA_fraction} shows that QA and QH have a larger stochastic fraction of barely-trapped particles than QI, due to the larger radial variance in magnetic field strength illustrated in figure~\ref{fig:harmonics_and_losses}. Despite the suppressed barely-trapped stochastic fraction in QI, figures~\ref{fig:harmonics_and_losses} and~\ref{fig:WBA_fraction} show significant stochastic prompt losses from trapped orbits in both QI and QH cases. Further work is needed to generalize the stochasticity onset criterion~(\ref{eq:stochasticity_threshold}) beyond passing orbits.
\section{Conclusion}
\label{sec:conclusion}
We extend the resonance analysis of~\citep{paul2023fast} to shear Alfv{\'e}n waves consistent with equilibrium field and Alfv{\'e}n velocity profiles, and show that these perturbations drive significant prompt alpha particle losses in optimized QA, QH, and QI stellarators at amplitudes $\delta\hat{B}_s \sim 10^{-3}$ consistent with observations in LHD~\citep{ogawa2012magnetic} and TFTR~\citep{hsu1992alpha}.

The resonance analysis for passing particles shows that increased field helicity $N$ widens the spacing between resonant islands in QH and QI but not QA, suppressing wave-induced stochastic transport in the former two cases. Notably, this suppression is not specific to SAW perturbations and applies to any transverse electromagnetic perturbation resonant with the passing particle drift motion. The passing resonance analysis also extends to barely trapped particles. We show that, because of reduced radial variation in magnetic field strength in QI, SAW-induced orbit-class transitions are suppressed in QI relative to QA and QH stellarators.

The resonance analysis is verified with kinetic Poincar{\'e} cross-sections and weighted Birkhoff averaging~\citep{duignan2023distinguishing, lachmann2025poincare} diagnostics implemented in the FIRM3D code~\citep{firm3d2025}. We note, however, that the resonance analysis does not account for radial variation in wave amplitude, which can independently structure transport as illustrated by the island-like features in the Poincar{\'e} cross-sections near $\Phi_{m,n}(s) = 0$. Radial amplitude variation is expected for SAWs, since Alfv{\'e}n velocity shear from plasma density profiles induces shear in the Alfv{\'e}n continuum and promotes continuum damping. This effect is particularly pronounced in QA case we considered, where narrow frequency gaps cause global modes to be fully damped under the $\rho_0^{\rm axis}(1-s^5)$ density shear, whereas radially global SAWs persist in QH and QI under the same shear due to wider gaps. This supports continuum gap optimization~\citep{Paul_Hyder_Rodriguez_Jorge_Knyazev_2025} as a viable approach to SAW-induced transport suppression. We also note that differences in harmonic composition of SAW eigenmodes across configurations complicate direct stochasticity comparisons between them.

Stochasticity analysis with WBA and loss fraction calculations show significant stochastic transport of trapped particles driven by SAW perturbations, motivating future extension of the resonance analysis beyond passing orbits. Although the presented analysis shows significant prompt alpha losses at amplitudes consistent with earlier experiments, determining SAW saturation amplitudes via quasilinear~\citep{gorelenkov2018resonance} or nonlinear models is needed for a self-consistent assessment of energetic particle confinement in stellarator fusion power plant designs.

\section*{\normalsize\textbf{Acknowledgements}}
We acknowledge funding through the U.S. Department of Energy, under contracts DE-SC0024630, DE-SC0024548 and DE-AC02-09CH11466. We also acknowledge funding through the Simons Foundation collaboration ``Hidden Symmetries and Fusion Energy,'' Grant No. 601958. This research used resources of the National Energy Research Scientific Computing Center (NERSC), a Department of Energy Office of Science User Facility using NERSC award ERCAP0031926.
\appendix

\section{Appendix: STELLGAP and AE3D harmonic resolution}
\label{sec:simulations}

We used STELLGAP and AE3D codes to calculate the Alfv{\'e}n continua and radially global shear Alfv{\'e}n waves for this study. Both codes represent a SAW as a superposition of cosine terms in Boozer angles. Such a representation is equivalent to (\ref{eq:deltaPhi}), and takes advantage of the fact that differential operators are applied an even number of times in continuum and vorticity~(\ref{eq:vorticity}) equations, preserving the cosine terms. This appendix describes the procedure used to determine appropriate dynamic resolution for the stellarators in Table~\ref{tab:equilibria}.

We choose dynamic harmonic numbers $(m,n)$ sufficient to resolve the shear Alfv{\'e}n continuum up to frequencies four times the Alfv{\'e}n frequency at the magnetic axis, $\omega \leq 4\omega_{\rm A}$. For a given dynamic harmonic, the continuum frequency satisfies the shear Alfv{\'e}n~\citep{Paul_Hyder_Rodriguez_Jorge_Knyazev_2025} dispersion relation $\omega/\omega_A = |\iota m - n|$. Therefore, to resolve $\omega \leq 4\omega_{\rm A}$, the poloidal harmonic range for each toroidal mode number $n$ is selected such that
\begin{align}
\label{eq:harmonic_resolution_criteria}
m_{\min} \leq m \leq m_{\max}, \quad \text{where} \quad |\iota m - n| \leq 4 \quad \text{for all} \quad \iota \in [\iota_{\min}, \iota_{\max}].
\end{align}
Notably, since STELLGAP and AE3D use a cosine basis, one should only include a single $(m,n)$ combination for each unique $\cos(m\theta-n\zeta)$ basis function. For example, for harmonics with $m=0$, we only include positive $n$ that match condition~(\ref{eq:harmonic_resolution_criteria}).

When the complex exponential~(\ref{eq:deltaPhi}) series is substituted into the vorticity~(\ref{eq:vorticity}) equation to find a SAW in a stellarator with $N_{\rm fp}$ field periods, the resulting expression factorizes into $N_{\rm fp}$ independent series called mode families. Because mode families are independent, they can be simulated separately. For the complex exponential series, harmonics with toroidal mode number $n_0$ will belong to the same mode family as harmonics with toroidal mode number equal to $n_0+jN_{\rm fp}$ for any integer $j$. Meanwhile, for the cosine series expansion, harmonics with toroidal mode number $n_0$ will belong to the same mode family as harmonics with both $n_0+jN_{\rm fp}$ and $-n_0+jN_{\rm fp}$ toroidal mode numbers, for any integer $j$.

Because STELLGAP and AE3D rely on forward and inverse Fourier transforms between the grid and harmonic representations, the equilibrium grid resolution must be sufficient to avoid aliasing errors. We use grid dimensions at least $5/2$ times the maximum harmonic numbers in each direction. The dynamic harmonic resolutions used for the stellarators from Table~\ref{tab:equilibria} are given in Table~\ref{tab:harmonics}.

\begin{table}
\centering
\begin{tabular}{cc|cc|cc}
\hline\hline
\multicolumn{2}{c|}{QA} & \multicolumn{2}{c|}{QH} & \multicolumn{2}{c}{QI} \\
$n$ & $m_{\min} \leq m \leq m_{\max}$ & $n$ & $m_{\min} \leq m \leq m_{\max}$ & $n$ & $m_{\min} \leq m \leq m_{\max}$ \\
\hline
 -2&  $1\leq m \leq11$  & -3 & $1\leq m \leq2$ & -3 & $1\leq m \leq2$\\
 0&  $1\leq m \leq20$   & -1 & $1\leq m \leq4$ & -1& $1\leq m \leq4$\\
 2&  $0\leq m \leq26$   &  1 & $0\leq m \leq6$ & 1&  $0\leq m \leq7$\\
 4&  $4\leq m \leq26$   &  3 & $0\leq m \leq8$ & 3&  $0\leq m \leq10$\\
 6&  $4\leq m \leq26$   &  5 & $0\leq m \leq10$ & 5&  $0\leq m \leq12$\\
 8&  $8\leq m \leq26$   &  7 & $2\leq m \leq12$ & 7&  $2\leq m \leq15$\\
10&  $12\leq m \leq26$  &  9 & $4\leq m \leq14$ & 9&  $4\leq m \leq17$\\
12&  $17\leq m \leq26$  &  11& $6\leq m \leq16$ & 11& $6\leq m \leq17$\\
14&  $20\leq m \leq26$  &  13& $7\leq m \leq18$ & 13& $8\leq m \leq17$\\
20&  $24\leq m \leq26$  &  15& $9\leq m \leq20$ & 15& $10\leq m \leq17$\\
  &                     &  17& $10\leq m \leq22$ & 17& $12\leq m \leq17$\\
  &                     &  19& $12\leq m \leq23$ & 19& $15\leq m \leq17$\\
  &                     &  21& $14\leq m \leq23$ &   & \\
  &                     &  23& $16\leq m \leq23$ &   & \\
  &                     &  25& $18\leq m \leq23$ &   & \\
  &                     &  27& $20\leq m \leq23$ &   & \\
  &                     &  29& $21\leq m \leq23$ &   & \\
\hline\hline
\end{tabular}
\caption{Dynamic Fourier harmonics $(m,n)$ used in STELLGAP and AE3D cosine series for stellarators from Table~\ref{tab:equilibria}.}
\label{tab:harmonics}
\end{table}

\section{Appendix: Resonance condition for QS equilibria}
\label{sec:appendix}
Let $\chi=M\theta-N\zeta$ be the helicity angle of a precisely quasi-symmetric $B_0(s,\theta,\zeta)=B_0(s,M\theta-N\zeta)$ equilibrium field. We choose $\chi$ to be a new angle coordinate, with a second coordinate $\eta$ determined by a general invertible linear transformation,
\begin{align}
\label{eq:harmonics_prime}
\theta=\frac{N_\eta\chi-N\eta}{MN_\eta-M_\eta N},\quad\zeta=\frac{M_\eta\chi-M\eta}{MN_\eta-M_\eta N},
\end{align}
where $MN_\eta-M_\eta N \neq 0.$ The unperturbed drift motion in these coordinates can be schematically expressed as 
\begin{align}
\begin{cases}
\label{eq:schematic_eoms}
\dot{\chi} &= \omega_\chi + \sum_{j\neq 0}\chi_j \cos(j\chi), \\
\dot{\eta} &= \omega_\eta + \sum_{j\neq 0}\xi_j \cos(j\chi),
\end{cases}
\end{align}
where $\omega_\chi=\langle\dot\chi\rangle$ and $\omega_\eta=\langle\dot\eta\rangle$ describe the time averaged drift $\langle A\rangle=\lim_{T\to\infty}\int_0^T dt A/T$, and $\chi_j$, $\xi_j$ describe periodic drift oscillations. Under assumption~(\ref{assump:singleharm}), the periodic drifts are dominated by $\nabla B_0$ drift from the largest Fourier mode of the equilibrium field, resulting in a single dominant harmonic $j=j'$ in unperturbed~(\ref{eq:schematic_eoms}) motion. Under the vacuum field~(\ref{assump:vacuum}) $I=0$ assumption, the radial displacement caused by a single Fourier harmonic $(m,n)$ SAW perturbation is 
\begin{align}
\label{eq:radial_perturbation_impact}
\delta\dot s=-\left(\omega +(\iota m -n)\right)\left(\frac{M_\eta\omega_\chi-M\omega_\eta}{MN_\eta-M_\eta N}+\frac{M_\eta\chi_{j'}-M\xi_{j'}}{MN_\eta-M_\eta N}\cos(j'\chi)\right)\frac{m}{\omega\psi_0}\Phi_{m,n}(s)\cos(\tau),
\end{align}
where $\tau=\omega t + m\theta-n\zeta =\omega t+m_\chi\chi-n_\eta\eta$ is the perturbation phase, and 
\begin{align}
m_\chi=\frac{mN_\eta-nM_\eta}{MN_\eta-M_\eta N},\;\quad n_\eta=\frac{mN-nM}{MN_\eta-M_\eta N}.
\end{align}
The phase along the unperturbed trajectory is 
\begin{align}
\tau=\tau^0+(\omega+m_\chi\omega_\chi-n_\eta\omega_\eta)t+\left(\frac{m_\chi\chi_{j'}}{j'\omega_\chi}-\frac{n_\eta\xi_{j'}}{j'\omega_\chi}\right)\sin(j'[\chi^0+\omega_\chi t]),
\end{align}
where $\tau^0=m_\chi\chi^0-n_\eta\eta^0$, $\chi^0=\chi(t=0),$ $\eta^0=\eta(t=0)$. Representing $\cos(\tau)$ as a series of Bessel functions $J_{k}$ of the first kind,
\begin{align}
\cos(\tau) = \sum_k J_k (\tau_{{j'}}) \cos \left(\tau^0 + \Omega_{kj'} t\right),
\end{align}
where $\tau_{j'}=(m_\chi\chi_{j'}-n_\eta\xi_{j'})/(j'\omega_\chi)$ and $\Omega_{kj'}=\omega+(m_\chi+kj')\omega_\chi-n_\eta\omega_\eta,$ allows one to express radial perturbation~(\ref{eq:radial_perturbation_impact}) as 
\begin{align}
\delta\dot s=&\sum_k s^0_{kj'}\cos(\tau^0+\Omega_{kj'}t)+ s^{-}_{kj'}\cos(\tau^0+j'\chi^0+\Omega_{kj'}t)+s_{kj'}^+\cos(\tau^0-j'\chi^0+\Omega_{kj'}t).
\end{align}
Here,
\begin{align}
\label{eq:drift_amplitudes}
s_{kj'}^0 &= -\frac{m}{\psi_0}\Phi_{m,n}(s) J_k(\tau_{j'})\left(1+(\iota m-n)\frac{M_\eta\omega_\chi/\omega-M\omega_\eta/\omega}{MN_\eta-M_\eta N}\right),\\
s_{kj'}^\pm &= -\frac{m}{2\psi_0}\Phi_{m,n}(s)J_{k\pm1}(\tau_{j'})\left(\iota m-n\right)\frac{M_\eta \chi_{j'}/\omega-M\xi_{j'}/\omega}{MN_\eta-M_\eta N}.
\end{align}
Naturally, the radial displacement amplitudes $s_{kj'}^\pm$ and $s_{kj'}^0$  are independent of the angle coordinate choices and match expressions from previous QH analysis~\citep{paul2023fast}, where the scaling of these coefficients with the drift magnitudes, mode numbers, and perturbed radial field was summarized as 
\begin{align}
\label{eq:scaling_amplitudes}
\begin{cases}
s^0_0 &\sim (\iota - h)J_0(\tau_{j'})\delta\hat{B}^s, \\ 
s^0_{kj'} &\sim J_k(\tau_{j'})\delta\hat{B}^s, \\ 
s^\pm_{kj'} &\sim J_{k\pm 1}(\tau_{j'})\xi_{j'}\delta\hat{B}^s 
\end{cases}
\end{align}
where $\delta \hat B^s$ is the normalized radial magnetic field~(\ref{eq:normalized_Bs}) perturbation and ${h=\omega_\theta/\omega_\zeta}$ is the helicity of the unperturbed orbit. The scaling analysis~\citep{paul2023fast} of radial displacement amplitudes~(\ref{eq:scaling_amplitudes}) shows that the $l=0$ direct wave interaction and the $l\pm1$ sideband resonances are the most significant for radial transport. Because there are several $l$ of significant displacement amplitude present for a single harmonic SAW, even a single harmonic SAW can cause island overlap and onset stochasticity.

\bibliographystyle{jpp}

\bibliography{lib}

\end{document}